\documentclass[]{JHEP3}

\usepackage{amsfonts}
\usepackage{amssymb}
\usepackage{latexsym}
\usepackage{indentfirst}
\usepackage{amsmath}

\newcommand{\g}{\gamma}
\def\d{\delta}
\newcommand{\D}{\Delta}

\def\eps{\varepsilon}

\def\k{\kappa}
\newcommand{\la}{\lambda}
\newcommand{\La}{\Lambda}

\newcommand{\si}{\sigma}
\newcommand{\Si}{\Sigma}

\def\cM{\mbox{${\cal M}$}}
\def\cN{\mbox{${\cal N}$}}
\def\cR{{\cal R}}
\def\cU{{\cal U}}

\def\cO{{\cal O}}

\newcommand{\gt}[1]{\mathfrak{#1}} 
\def\dim{{\rm dim}}

\def\det{{\rm det}}

\def\tr{{\rm tr}}

\newcommand{\rea}[1]{\tx{REA}_q(#1)}
\def\emb{\hookrightarrow}

\def\llap#1{\hbox to 0pt{\hss#1}}
\def\pola{a\llap{\hbox{\char'30\kern-1.2pt}}}
\def\pole{e\llap{\hbox{\char'30\kern-.8pt}}}

\newcommand{\non}{\nonumber\\}

\newcommand{\p}{\partial}

\def\x{\times}
\def\ox{\otimes}

\def\too{\longrightarrow}

\newcommand{\lb}{\left(}
\newcommand{\rb}{\right)}

\newcommand{\lan}{\langle}
\newcommand{\ran}{\rangle}

\newcommand{\beq}{\begin{equation}}
\newcommand{\eeq}{\end{equation}}
\newcommand{\beqa}{\begin{eqnarray}}
\newcommand{\eeqa}{\end{eqnarray}}
\newcommand{\barr}{\begin{array}}
\newcommand{\earr}{\end{array}}
\newcommand{\ben}{\begin{enumerate}}
\newcommand{\een}{\end{enumerate}}
\newcommand{\bit}{\begin{itemize}}
\newcommand{\eit}{\end{itemize}}

\newcommand{\refeq}[1]{(\ref{#1})}

\newcommand{\rep}{representation }
\newcommand{\repn}{representation}
\newcommand{\reps}{representations }

\newcommand{\irrep}{irreducible representation }

\newcommand{\irreps}{irreducible representations }

\newcommand{\hw}{highest-weight }

\newcommand{\rhs}{the right-hand side }

\newcommand{\tx}[1]{\textrm{#1}}

\def\trr{\triangleright}

\newcommand{\id}{{\rm id}}
\newcommand{\End}{{\rm End}}

\def\v{{\rm v}}

\def\mg{\gt{g}}

\def\mgh{\widehat{\gt{g}}}

\def\gcheck{g^{\vee}}

\def\gcg{\cU_q(\mg^L \x \mg^R)_\cR}

\def\uqg{\cU_q(\mg)}

\newcommand{\faff}[1]{P^{\k}_{+}(#1)}

\def\Mo{M^{(0)}}

\def\1{\mathbb{I}}
\def\spun{\tx{span}}
\newcommand{\matR}[4]{R_{#1 #2 , #3 #4}}

\newcommand{\lhs}{the left-hand side }

\newenvironment{princ}{~\newline\newline\textbf{Principle.}}{~\newline\newline}
\newenvironment{claim}{~\newline\newline\textbf{Claim.}}{~\newline\newline}
\newenvironment{conc}{~\newline\newline\textbf{Conclusion.}}{~\newline}

\title{Brane Bulk Couplings and Condensation from REA 
Fusion\footnote{Work supported by Polish State Committee 
for Scientific Research (KBN) under contract 2 P03B 001 
25 (2003-2005). R.R.S. was additionally supported by the Marie Curie
Fellowship at King's College London.}}\author{Jacek Pawe\l czyk$^a$
  and Rafa\l ~R. Suszek$^{ab}$\thanks{Current address: Laboratoire
    de Physique, \'Ecole Normal Sup\'erieure de Lyon; 46, all\'ee
    d'Italie, 69364 Lyon, France.}\\    
\normalsize{$^a$ \textit{Institute of Theoretical Physics, 
Warsaw University, \\ \ \ ul. Ho\.za 69, PL-00-681 Warsaw, Poland}} \\
\normalsize{$^b$ \textit{King's College London \\
\ \ Strand, London WC2R 2LS, U.K.}} \vspace{+0.5cm}\\
E-mail: \email{Jacek.Pawelczyk@fuw.edu.pl, 
rafal.suszek@ens-lyon.fr}}

\abstract{The physical meaning of the Reflection Equation Algebras 
of \cite{qalg} is elucidated in the context of Wess--Zumino--Witten 
D-brane geometry, as determined by couplings of closed-string modes 
to the D-brane. Particular emphasis is laid on the r\^ole of algebraic 
fusion of the matrix generators of the Reflection Equation Algebras. 
The fusion is shown to induce transitions among D-brane configurations 
admitting an interpretation in terms of RG-driven condensation phenomena.}


\begin{document}

\section{Introduction}\label{intro}

Physics of D-branes has long been a subject of intense study. Of
particular interest are D-branes on compact WZW manifolds. They provide
interesting examples of D-brane behaviour in non-trivial backgrounds
with fluxes. But even for these, highly symmetric, cases the full BCFT
analysis is rather complicated \cite{schom-lect} and stands in
shocking contrast with the very simple matrix model of D-brane
condensation advocated in \cite{fuzzy-ars}. There is a price to pay
for the simplicity of the latter --- it cannot describe all D-branes on
the group manifold. Some years ago, a matrix model based on
quantum-group symmetries was advanced \cite{qalg}. It seems to work
properly for all D-branes but many of its features are still
mysterious. The model uses the celebrated Reflection Equation (RE)
\cite{re} and its representation theory to derive D-brane properties. On
the one hand, the RE encodes a quantum-group version of the familiar
pattern of bulk symmetry breakdown of the underlying WZW model
resulting from the introduction of a maximally symmetric boundary ---
this is the particular aspect of it emphasised in \cite{kss} (the
quantum group of interest is the Drinfel'd--Jimbo deformation of the 
universal enveloping algebra, suggested by certain fundamental
structures of \linebreak the associated CFT). On the other hand, it defines
a quantisation of the distinguished Poisson structure on the target Lie
group of the WZW model, first elucidated by Semenov-Tian-Shansky in 
\cite{sts}, compatible with the foliation of the group manifold by
conjugacy classes (to be wrapped by the maximally symmetric D-branes)
and forming part of \linebreak the canonical structure of the boundary WZW
model itself\footnote{We refer the reader to \cite{gaw} for a lucid 
  exposition and comprehensive bibliography.}.  

In this paper, we shall follow this track and analyse fusing
properties of the matrices entering the RE, which - as it turns out -
shed some light on the physical content of the algebra. It appears
that there are two types of fusion. We shall show how both of them
lead to some known D-brane properties. As we shall see, the first type
of fusion (which we dub the Bound-State Fusion -- the BSF) can be
interpreted as describing a process of formation of extended D-branes
out of D$0$-branes. It also chooses a particular set of
representations of the RE as the physically relevant ones. The second
type (to be called the Bulk-Weight Fusion -- the BWF) is just the
standard representation-theoretic fusion of the function algebra on a
given D-brane. Mastering this last kind of fusion is necessary to keep
control of functions on a given D-brane and yields couplings of gravity
to the D-brane. Our study thus takes us one step beyond the purely
geometric framework developed heretofore and enables us to make
contact with the rich stringy physics of \linebreak the parent WZW models.

\section{A lightning review of the algebraic setup}\label{sec:review} 

We begin by recapitulating the essential aspects of the quantum matrix 
models studied in \cite{qalg,qdyn,qorb}. The reader is urged to
consult the original papers for details. 

The central element of the quantum-group-covariant approach to the
study of non-classical D-brane geometry in compact (simple-)Lie-group  
targets is the Reflection Equation (RE) \cite{re}: 
\beq\label{re}
R_{21}^{\La_1 , \La_2} M^{\La_1}_1 R_{12}^{\La_1 , \La_2} 
M^{\La_2}_2 = M^{\La_2}_2 R_{21}^{\La_1 , \La_2} 
M^{\La_1}_1 R_{12}^{\La_1 , \La_2},
\eeq
written for operator-valued matrices\footnote{Displaying the indices
  explicitly, the RE reads (here, $M_{ij} \equiv M^\La_{ij}$)    
$$ 
({\rm RE })_{ij,kl}: \qquad \matR kcia \ M^{\La_1}_{ab} \
  \matR bjcd\ M^{\La_2}_{dl} = M^{\La_2}_{kc} \ \matR cdia \
  M^{\La_1}_{ab} \ \matR bjdl.
$$
The indices $\{i,j,a,b\}$ and $\{k,l,c,d\}$ correspond to the first ($1$) and
the second ($2$) vector space in \refeq{re}, respectively.}
$M^{\La_{1,2}} \in \End(V_{\La_{1,2}}) \ox \rea{\mg}$, with
$\rea{\mg}$ \linebreak the (abstract) Reflection Equation Algebra (REA) and
$V_{\La_{1,2}}$ two irreducible modules of $\uqg$, labeled by the
respective highest weights $\La_1, \La_2$. Here, the deformation
parameter is $q = e^{\frac{\pi i}{\k + \gcheck(\mg)}}$, a value
suggested by a detailed analysis of the (B)CFT of the WZW models of
interest, and $R_{12}^{\La_1,\La_2} = (\pi_{\La_1} \ox
\pi_{\La_2})(\cR_{12})$ is the suitably
represented\footnote{$\pi_{\La}$ is the \irrep of $\uqg$ of the
  highest weight $\La$, a dominant integral affine weight of $\mg$. We
  denote the set of all such weights (the fundamental affine alcove)
  by $\faff{\mg}$. In particular $\pi_{\La_F}$ stands for the
  fundamental (defining) \repn.} $\cR$-matrix of the Drinfel'd--Jimbo
quantum group $\uqg$. The RE \refeq{re} is readily verified to induce
a $\gcg$-module structure on $\rea{\mg}$ under which $M^\La$ transform
as elements of the tensor module $V^{(L)}_\La \ox V^{(R)}_{\La^+}$
(here, $\gcg$ is just $\cU_q(\mg^L) \ox \cU_q(\mg^R)$ as an algebra,
with a suitably twisted coalgebra structure \cite{qalg}) - indeed, the
bichiral transformations:     
\beq\label{UU-action}
(u_L \ox u_R) \trr M^\La_{ij} = \pi_\La(Su_L)_{ik} M^\La_{kl}
\pi_\La(u_R)_{lj},  
\eeq
with $S$ the antipode of $\uqg$, preserve \refeq{re} (i.e. transform
solutions into solutions). \linebreak The left-right symmetry of the RE is
to be regarded as a quantum counterpart of \linebreak the left-right $\mg^L
\ox \mg^R$-symmetry of the target group manifold, the horizontal component
of the (Kac--Moody) $\mgh_\k^L \ox \mgh_\k^R$-symmetry of the WZW model in
the bulk.  

Given the REA defined by the above commutation relations, together
with the additional quantum-determinant constraint ($M \equiv
M^{\La_F}$):   
\beq\label{qdet}
\det_q M \overset{!}{\propto} 1,
\eeq
to be interpreted as fixing the ``size'' of the quantum group
manifold, we may subsequently consider its \irreps
separated\footnote{Cf \cite{qorb} for a discussion of (geometrically  
  well-understood) degeneracies.} by the Casimir operators:
\beq\label{cas}
\gt{c}_k^\La = (\tr_q \ox \id)(M^\La)^k.
\eeq
Upon descending to any specific such \repn, we break the original
left-right symmetry down to the diagonal part:
\beq
\gcg \ni u_L \ox u_R \searrow u \ox u \in (\gcg)^V \cong \cU_q(\mg) 
\eeq
which fits in well with the picture of reduction of the bulk symmetry
at an untwisted maximally symmetric boundary\footnote{For an extension
to the twisted case, see: \cite{qtwist}.}. This elementary observation
already hints at the viable identification of \irreps of the REA with
(untwisted) maximally symmetric boundary conditions of the relevant
WZW model, that is with (untwisted) maximally symmetric D-branes.

In order to give some flesh to the last statement, we need an explicit
realisation of \linebreak the defining relations \refeq{re} and
\refeq{qdet}. Luckily, one particular such realisation has long been 
known \cite{kss,dkm} ($M_0^\La$ is an arbitrary $c$-number solution to
the RE): 
\beq\label{frtemb}
M^\La = L^+ M_0^\La SL^- \equiv (\pi_\La \ox \id) (\cR_{21}) M_0^\La
(\pi_\La \ox \id) (\cR_{12})    
\eeq 
and is determined by the Faddeev--Reshetikhin--Takhtajan (FRT)
realisation \cite{frt}:
\beq
L^{+}_{ij} = [(\pi_\La)_{ij} \ox \id] (\cR_{21}) \qquad , \qquad 
L^{-}_{ji} = [(\pi_\La)_{ij} \ox \id] (\cR_{12}^{-1}), \qquad 1
\leq i \leq j \leq \dim V_\La 
\eeq 
of the Drinfel'd--Jimbo quantum group $\uqg$ in terms of the so-called
$L^\pm$-operators satisfying the standard $R$-matrix commutation
relations. \emph{Choosing} this realisation for \linebreak the REA renders at our disposal the well-developed \rep theory of $\uqg$ whose  peculiarities
at the CFT-dictated root-of-unity value of the deformation parameter $q$
have provided ample evidence for an intimate relationship between the REA
thus reconstructed and quantum D-branes, as summarised below:  
\bit
\item[-] \irreps $\pi_\La$ of $\uqg$ of a non-vanishing quantum
  dimension ($\La \in \faff{\mg}$) are in one-to-one correspondence
  with the inequivalent untwisted maximally symmetric boundary states
  $|\La \rangle\rangle_C$ of the WZW model (a twisted variant of the
  correspondence has been worked out in \cite{qtwist}), the truncated
  tensor product of \linebreak these \reps reproduces the fusion rules of
  the latter - cf \cite{qalg}; 
\item[-] the \rep theory of the REA induced from that of the quantum
  group accounts well for the discrete symmetries of the group
  manifold generated by the simple currents of the CFT - cf
  \cite{qorb};  
\item[-] harmonic analysis on the quantum geometries associated with
  the \irreps of the REA agrees with the decoupling limit
  \cite{fuzzy-ars,geowzw} of the subalgebra of \linebreak the boundary OPE
  algebra composed of horizontal multiplets descended from primary boundary
  fields that do not change the boundary condition - cf \cite{qalg}; 
\item[-] localisation of D-branes within the quantum group manifold
  from fixing Casimir eigenvalues is in keeping with the semiclassical 
  results - cf \cite{qalg,qtwist};
\item[-] exact values of D-brane tensions follow from a general
  matrix-action \textit{Ansatz} - cf \cite{qalg};
\item[-] a well-defined semiclassical limit coincides with the
  perturbative fuzzy structure of \cite{fuzzy-ars} - cf \cite{qalg};
\item[-] fractionation of D-branes at fixed-points of simple-current
  orbifold action admits a straightforward  algebraic description - cf
  \cite{qorb}.  
\eit
In this paper, we attempt to give an independent justification of the
choice of realisation of the REA that underlies the hitherto
  successful quantum reconstruction programme, whereby we also
  discover an algebraic description of the D-brane condensation
  phenomena responsible for creation of arbitrary D-branes of the
  model (of the kind described) from gauge-field-perturbed stacks of
  elementary D$0$-branes. Last, we rederive harmonic analysis on any
  given D-brane from the analysis of an algebraic fusion procedure and
  - most importantly - extract from the FRT-realised REA the
  microscopic D-brane geometry data, as encoded in the graviton
  coupling to its worldvolume.

\section{Bound-state fusion}\label{sec:BSF}

In this section, we establish a non-trivial link between effective
D-brane gauge dynamics in boundary WZW models and the REA's defined by
\refeq{re}. We want to introduce an algebraic cousin of the D-brane
condensation effect \cite{myers}, discussed at great length in
\cite{fuzzy-ars,condo} with reference to the seminal papers by Affleck
and Ludwig \cite{aff-lud}. In the case at hand, the very form of
fusion leads us to conclude that an arbitrary D-brane, as described by
its function algebra, can be built out of a number of trivial
representations of the RE, describing D$0$-branes. The algebraic
fusion algorithm has been devised in direct reference to the
techniques of the principal chiral model presented in \cite{mackay},
in which there is additional structure (dependence on a dynamical
parameter) justifying its interpretation. In the present setup,
lacking this extra structure, some elementary tests of its validity
are performed explicitly below, as well as in the Appendix. In
particular, we verify - rather importantly - that it has the expected
semiclassical limit.  
 
We begin by remarking that the operator-valued
matrix\footnote{Here, we add an extra index $(\La_B)$ which indicates
  that $M^{\La(\La_B)} \in \End(V_{\La}) \ox \End(V_{\La_B}) \ox
  \rea{\mg}$. Earlier analyses focused mainly on the special case
  $(\La,\La_B) = (\La_F,0)$ in which $M^{\La_F(0)} \equiv M^{\La_F}$
  represents the coordinate module of the quantised group
  manifold. The meaning of $M^\La$ for a general $\La$ shall be
  expounded in the next section.} $M^{\La(\La_B)}$ given by either
side of \refeq{re},         
\beq\label{BSF-0}
M^{\La(\La_B)}_1 = M^{\La_B}_2 R_{21}^{\La,\La_B} M^{\La}_1
R_{12}^{\La,\La_B},  
\eeq
also satisfies an appropriate RE:
\beq\label{BSF-RE}
R_{21}^{\La_1 , \La_2} M^{\La_1(\La_B)}_1 R_{12}^{\La_1 , \La_2}
  M^{\La_2(\La_B)}_2 =  M^{\La_2(\La_B)}_2 R_{21}^{\La_1 , \La_2}
  M^{\La_1(\La_B)}_1 R_{12}^{\La_1 , \La_2}
\eeq
for $\La_B$ arbitrary. The latter follows straightforwardly from the 
RE's and the Quantum Yang--Baxter Equation satisfied by the
$M$-matrices fused and the $\cR$-matrix, respectively. In other words,
the Bound-State Fusion (BSF) thus defined, \refeq{BSF-0}, provides a
systematic method of generating new solutions to the RE from the known 
ones. 

\def\Mo{(M_0)}
The physical significance of \refeq{BSF-0} relies on the observation
that it singles out a set of REA representations of special relevance
to the study of WZW D-branes. Take any $c$-number matrix $\Mo^\La$
respecting the RE (considered, e.g., in \cite{kss}) so that
$\Mo^{\La_B}_2 \Mo^{\La}_1$ also satisfies the RE. According to the
logic outlined in Sec.\ref{sec:review}, the latter is - for
$\Mo^{\La_B} = \1$ (the unit matrix of dimension $\dim V_{\La_B}$) -
to be associated with $\dim V_{\La_B}$ D$0$-branes located at
positions defined by $\Mo^\La$ as \textit{per} Casimir eigenvalues
\refeq{cas}. Then, \rhs of \refeq{BSF-0} belongs to $(\pi_\La \ox
\pi_{\La_B})(\uqg \ox \uqg)$. We interpret the process of passing from
the reducible representation just described, $\Mo^{\La_B}_2
\Mo^{\La}_1$, to the irreducible one given by \refeq{BSF-0} as
condensation and depict it as     
\beq\label{a-br}
\Mo^{\La_B}_2 \Mo^{\La}_1\ \too\ M^{\La(\La_B)} \equiv
R_{21}^{\La,\La_B} \Mo^{\La}_1 R_{12}^{\La,\La_B}.   
\eeq
This, however, is none other but the FRT realisation \refeq{frtemb} of 
the \irrep of $\rea{\mg} \emb \uqg$ of highest weight $\La_B$,
\emph{chosen} in \cite{qalg} for the simple reason: it induces a \rep 
theory of $\rea{\mg}$ whose elements, irreducible \hw \reps of $\uqg$
of a non-vanishing quantum dimension, are in a straightforward
one-to-one correspondence with all the candidate algebraic D-branes
associated, in \cite{geowzw,geowzwas}, with untwisted maximally
symmetric WZW boundary conditions on the compact (simple and simply
connected) Lie group $G$. Thus, we can postulate the following 
\begin{princ} 
\emph{Untwisted maximally symmetric quantum WZW D-branes on a simple
  and simply connected compact Lie group $G$ are classified by those
  \irreps of $\rea{\mg}$ which can be generated through the
  Bound-State Fusion \refeq{BSF-0} from an elementary $c$-number
  D$0$-brane solution.}
\end{princ}  
The BCFT-preferred FRT realisation is now an immediate consequence of
the Principle whose physical rationale shall be presented below. 

We may next apply the fusion algorithm to the physical solutions
generated from \linebreak the D$0$-brane one. Thus, given $M^{\La_B(\la)}$
and $M^{\La(\la)}$ the fusion \refeq{BSF-0} leads to:   
\beq\label{BSF}
M^{\La(\La_B \x \la)} = M^{\La_B(\la)}_2 R_{21}^{\La,\La_B}
M^{\La(\la)}_1 R_{12}^{\La,\La_B}. 
\eeq
Here, \lhs belongs to $[\pi_\La\ox (\pi_{\La_B}\ox \pi_\la)](\uqg \ox
\uqg \ox \uqg)$ and hence - as a tensor operator\footnote{Cp
  \cite{qalg} and the Appendix.} - it can be decomposed as    
\beq\label{BSF-decomp}
M^{\La (\La_B \x \la)} = \oplus_{\mu \in \faff{\mg}} \cN_{\La_B \;
  \la}^{\quad \ \mu} M^{\La ( \mu)}, 
\eeq 
where $\cN_{\La_B \; \la}^{\quad \ \mu}$ are the standard fusion 
rules of the WZW model with the current symmetry $\mgh_\k$ and the
usual restriction to \irreps of $\uqg$ of a non-vanishing quantum
dimension has been imposed. 

The last result as well as the reasoning
that has led us to formulate the Principle are strongly reminiscent of 
the BCFT picture in which gauge-field perturbations induce transitions
through condensation between an original stack of D-branes and a final
(metastable) state. Let us dwell on this a little longer.

The fusion operation:
\beq
M^{\La(\la)} \xrightarrow{\La_B-BSF} M^{\La(\La_B \x \la)}
\eeq
defined above mimics the BCFT transition
\cite{condo}: 
\beq\label{kfin-cond}
(\la;\dim \La_B) \xrightarrow{A^{\La_B}} \oplus_{\mu \in \faff{\mg}}
\cN_{\La_B \; \la}^{\quad \ \mu} (\mu;1),
\eeq 
of a stacked $\dim V_{\La_B}$-tuple of D-branes of weight label $\la$,
effected by the marginal perturbation: $\int_{\p \Si} dt A^{\La_B}_a
J^a(t)$ ($\p \Si$ is the boundary of the open-string worldsheet) of
the boundary WZW model coupling the \emph{constant} gauge field
$A^{\La_B}_a = \pi_{\La_B}(T_a) \ox \1_{d_\la}${}\footnote{$\pi_{\La_B}(T_a)$
are the generators of the horizontal Lie algebra $\mg$, satisfying
the defining relation $[T_a,T_b] = \imath f_{abc} T_c$, and realised in the
\rep $\pi_{\La_B}$; moreover, we have denoted $d_\la := \dim V_\la$.} to
the boundary symmetry current $J$. In the  relevant fuzzy matrix model 
\cite{fuzzy-ars}, the transition is realised by perturbing the
background geometry $Y_a^{\la,d_{\La_B}} = \1_{d_{\La_B}} \ox
\pi_\la(T_a)$ of a stack of ($d_{\La_B}$) fuzzy D-branes of weight
label $\la$ with the specific gauge fluctuation $A^{\La_B}_a$ as   
\beq\label{kinf-cond}
Y_a^{\la,d_{\La_B}} \xrightarrow{A^{\La_B}} Y_a^{\la,d_{\La_B}} +
A^{\La_B}_a = \oplus_{\mu \in P_+(\mg)} L_{\La_B \; \la}^{\quad \ \mu}\
Y_a^{\mu,1},  
\eeq
whereby a semiclassical (large-$\k$) variant of the condensation
effect is induced ($L_{\La_B \; \la}^{\quad \ \mu}$ are the
Littlewood--Richardson coefficients which replace the fusion rules at
large values of \linebreak the level). Motivated thus, we put forward the
following 
\begin{claim} 
\emph{The Bound-State Fusion \refeq{BSF} captures - in the algebraic
  framework of the REA - the gauge-field-driven effect of condensation 
  with boundary-spin absorption \refeq{kfin-cond}.} 
\end{claim} 
In order to substantiate it, we need to go back to \cite{qdyn} and
identify nontrivial gauge-field degrees of freedom on a stack of
quantum D-branes. Hence, we associate small (we have \linebreak a natural
expansion parameter $\hbar \equiv \frac{\pi}{\k +\gcheck(\mg)}$)
gauge-field excitations -  in the vein of a much more general approach
to gauge fields on a noncommutative geometry - with perturbations of
the geometric background, $M^{\La(\la)}, \ \La = \La_F$ (the
coordinate module), exactly as in the semiclassical picture
\refeq{kinf-cond}. Furthermore, we decompose some of the terms in
\refeq{BSF}, $X \in
\{M^{\La_B(\la)}_2,R_{21}^{\La_F,\La_B},R_{12}^{\La_F,\La_B}\}$, as $X
= \1 \ox \1 + x$, where $x$ is of the order of $\cO(\hbar)$ and
($q$-)traceless (up to corrections of higher order in $\hbar$). With
this decomposition, in which we assume\footnote{This is precisely the
  domain of validity of the semiclassical approximation.} that $0
\lesssim \Vert \la \Vert, \Vert \La_B \Vert \ll \k$, the leading term
in \refeq{BSF} reads $\si_{1,2}(\1_{d_{\La_B}} \ox M^{\La_F(\la)})$
($\si_{1,2}$ interchanges the first and second tensor components) and
shall be denoted by $M^{\La_F(\La_B\x\la)}_0$. Thus, \rhs of
\refeq{BSF} can be rewriten as     
\beq
M^{\La_F(\La_B \x \la)} = M^{\La_F(\La_B\x\la)}_0 +
A^{\La_F(\La_B\x\la)}, 
\eeq
where $A^{\La_F(\La_B\x\la)}\sim \cO(\hbar)$ acquires the
interpretation of a gauge field\footnote{Using the BWF of
  Sec.\ref{sec:to-grav}, the gauge field is readily shown to be a
  ``function'' of the background geometry $M^{\La_F(\la)}$.}. Precise
agreement between our description\footnote{It is worth remarking that
  the other natural (given the FRT realisation) candidate for an
  algebraic description of the condensation phenomena, namely the
  standard coproduct in the second tensor component of $\cM = \cR_{21}
  \cR_{12}$ in \eqref{frtemb}, yields essentially the same result.}
and the BCFT one \eqref{kinf-cond} follows from the fact that the RE 
\eqref{BSF-RE}, satisfied by $M^{\La_F(\La_B \x \la)}$, reproduces -
in the semiclassical r\'egime, at $\cO(\hbar)$ - exactly the matrix 
equations of the fuzzy model of the BCFT \cite{qalg} satisfied by \rhs
of \eqref{kinf-cond}, that is the vanishing-curvature equation for the
gauge potential $A^{\La_B}$ (in \linebreak this picture, the semiclassical
transformation rules for $A^{\La_B}$ become a consequence of \linebreak those
of the covariant coordinate $M^{\La_F(\La_B \x \la)}$). Alternatively, 
one may perform an $\hbar$-expansion of the explicit FRT realisation
of $M^{\La_F(\La_B \x \la)}$, whereby one readily reobtains 
\eqref{kinf-cond} at the first nontrivial level.

\section{Bulk-weight fusion and brane-gravity couplings}\label{sec:to-grav}

Let us begin by recalling that the quantised algebra of functions on
untwisted D-branes, $\rea{\mg}$, is generated by the elements
$M^{\La_F}_{ij}$. There is a natural basis of \linebreak the algebra, regarded
here as a vector space, namely the basis of $\uqg$-intertwiners
related directly - in the physical context - to the multiplets of
boundary fields on a given D-brane descended from the primary fields
of the BCFT by the action of the horizontal subalgebra $\mg$ of the
current symmetry algebra $\mgh_\k$ of the relevant WZW model (in \linebreak
the decoupling limit of \cite{fuzzy-ars,geowzw}). We claim that       
\beq\label{basis}
\rea{\mg} = \oplus_{\La \in \faff{\mg}} \spun \langle M^\La_{ij}
\rangle_{i,j \in \overline{1,d_{\La}}},
\eeq
where $M^\La_{ij}$ is the $(i,j)$-th operator entry of the matrix
$M^\La$, is the basis sought. Above, $M^\La$ denote matrices
respecting the RE \refeq{re} written in the representation $\pi_\La 
\ox \pi_\La$. Thus, $M^{\La(\la)}$ (see the previous section) are -
indeed - quantum-group tensors with transformation properties
appropriate for functions on the standard set of D-branes, 
\beq\label{comm-proof}
(1\ox \pi_\la (u_1)) M^{\La(\la)} (1 \ox \pi_\la (Su_2)) = 
(\pi_\La (Su_1) \ox 1) M^{\La(\la)} (\pi_\La (u_2) \ox 1).
\eeq
The road to \refeq{basis} goes through the definition of the
Bulk-Weight Fusion (BWF): 
\beq\label{BWF}
M_{12}^{\La_1 \x \La_2} = \lb R_{12}^{\La_1,\La_2} \rb^{-1}
M^{\La_1}_1 R_{12}^{\La_1,\La_2} M^{\La_2}_2. 
\eeq
The fusion is a solution-generating operation for the ``bulk''
(matrix) tensor component of $M^{\La(\la)}$, compatible with the
defining relation \refeq{re} of a $\gcg$-module \cite{dkm}. It is \linebreak
a natural counterpart of the classical tensoring procedure\footnote{It is,
in particular, equivalent to the standard coproduct in the first tensor component
of the universal $M$-matrix $\cM = \cR_{21} \cR_{12}$ in the FRT realisation.}
in the category of solutions to the RE - one can easily show that \refeq{BWF}
yields a $(\pi_{\La_1}\ox\pi_{\La_2})(\gcg \ox \gcg)$-module and respects
the corresponding RE\footnote{The RE in question is \refeq{re} with both
$\pi_{\La_1}$ and $\pi_{\La_2}$ replaced by $\pi_{\La_1} \ox \pi_{\La_2}$.};
the right-hand side of \refeq{BWF} can be projected onto irreducible components,
$M^\La$, with $\pi_\La \subset \pi_{\La_1} \ox \pi_{\La_2}$. Thus, starting
from $M^{\La_F}$ ($\pi_{\La_F}$ is the defining representation of $\uqg$)
we can generate a basis of matrices $M^{\La}$ for \linebreak any $\La\in
\faff{\mg}$. This leads directly to \refeq{basis}.     

The above iterative algorithm for obtaining tensor-product solutions
from some given elementary ones, $M^{\La_1}$ and $M^{\La_2}$, is our
second example of RE fusion and was discussed at great length, in the 
above form, in \cite{dkm}. As we already know, it is not the only way
of composing elements of $\rea{\mg}$. We shall therefore distinguish
it by giving it a name suggested (once more) by the literature on the 
$(1+1)$-dimensional models, that is \linebreak the Bulk-Weight Fusion. 

Below, we give an interpretation to $\pi_\la(M^\La_{ij})$. Recall that
$\pi_\la(M^\La_{ij}) \in \End(V_\la)$ for \linebreak the D-brane labeled
by the weight $\la$. Accordingly, we may calculate the ($q$-)trace of $M^{\La}$
over the module $V_\la$. It is straightforward to demonstrate \cite{qalg}
that the trace is proportional to the unit matrix, that is  
\beq\label{si-tr}
\tr_q^{(\la)} (M^{\La}_{ij}) = \tr_{V_\la} (M^{\La}_{ij} \cdot
q^{2H_\rho}) = f(\La,\la)\ \d_{ij},  
\eeq
where $\rho$ is the Weyl vector of $\mg$. As shown in \cite{qalg},
$\pi_\la(M^\La_{ij})$ in the FRT realisation \refeq{a-br} encode a
number of properties of the weight-$\la$ D-brane in an algebraic manner
(cp Sec.\ref{intro}). Since we have not normalised $M$ so far, we can 
specify the function $f(\La,\la)$ up to a $\la$-dependent factor
only. Let us calculate $f(\La,\la)$. We use \cite{klimsch}     
\beq\label{r-expl}
\cR_{12}= q^{H_i F_{ij} \ox H_j} \lb I \ox I + \sum_{U^\pm} U^+ \ox
U^- \rb,  
\eeq
while for $\cR_{21}$ we transpose $U^+ \leftrightarrow U^-$ 
in the expression above. Here, $F$ is the (symmetric) quadratic matrix 
of $\mg$, and $U^+, U^-$ stand for terms in the Borel subalgebras 
of rising and lowering operators, respectively. As \lhs of
\refeq{si-tr} does not depend on \linebreak the vector from the module $V_{\La}$
that it acts upon, we can evaluate it on the highest-weight vector
(annihilated by $U^+$), $| \La \ran$. Then, only the generators of the 
Cartan subalgebra in \refeq{r-expl} contribute, 
\beq
q^{2H_i F_{ij} \ox H_j}\vert_{| \La \ran \ox \cdot} = q^{2 \La_i  
F_{ij} \ox H_j} = I \ox q^{2 H_{\La}},
\eeq
so that \refeq{si-tr} becomes\footnote{The explicit formula relating
  entries of the modular $S$-matrix to Lie-algebra characters can be
  found, e.g., in \cite{cftbook}.}
\beq\label{explicit-f}
\tr_q^{(\la)} (M^{\La}_{ij}) = \d_{ij} \ \tr_{V_\la}(q^{2(H_{\La} +
  H_\rho)}) = \d_{ij} \ \chi_\la \lb \frac{2 \pi i (\La + \rho)}{\k +   
\gcheck(\mg)} \rb = \d_{ij} \ \frac{S_{\La^+ \la}}{S_{\La^+ 0}},
\eeq
where $\chi_\la$ and $S_{\La^+ \la}$ are the standard character over
the $\mg$-module of highest weight $\la$ and the modular matrix of the
WZW model associated to $\mgh_{\k}$, respectively, whereas $\La^+$ is
\linebreak the unique charge conjugate of the weight $\La$. In particular,
for $\mg = \gt{su}_2$, we obtain     
\beq
f_{\gt{su}_2}(\La,\la) = \tr_{V_\la} (q^{(\La+1)H}) = 
\frac{\sin \frac{\pi(\La+1)(\la+1)}{k+2}}{\sin 
\frac{\pi(\La+1)}{k+2}},
\eeq 
which agrees with $S^{\gt{su}_2}_{\La \la} = \sqrt{\frac2{k+2}}\sin  
\frac{\pi(\La+1)(\la+1)}{k+2}$ and $\La^+ \equiv \La$ for all $\La$.

In order to understand the physics behind the last result, recall that
- on the BCFT side - untwisted maximally symmetric D-branes of the WZW
model are represented by Cardy states \cite{cardy}:  
\beq\label{BSCardy}
| \la \ran\ran_C= \sum_{\La \in \faff{\mg}} \frac{S_{\La
    \la}}{\sqrt{S_{\La 0}}}\ | \La \ran\ran_I. 
\eeq
Above, $| \La \ran\ran_I$ are the Ishibashi (character) states 
\cite{ishibcft}. The data encoded in \refeq{BSCardy} turn out to be 
sufficient to determine, to the leading order in $\hbar$, the coupling
of {\bf graviton} modes: 
\beq
| a,b,\g_{ij} \ran = J_{-1}^{(a} \tilde{J}_{-1}^{b)} | \g_i \ran \ox |   
\g^+_j \ran, \qquad | \g_i \ran \ox | \g^+_j \ran \in \widehat{V}_\g
\ox \widehat{V}_{\g^+} 
\eeq
to the D-brane defined by \refeq{BSCardy} (here, $J_{-1}$ and
$\tilde{J}_{-1}$ are the $(-1)$-th Laurent modes of \linebreak the two chiral
components of the bulk $\mgh_\k^L \ox \mgh_\k^R$-symmetry current,
acting on the $\mgh_\k$-modules $\widehat{V}_{\g,\g^+}$ of highest
weights $\g,\g^+ \in \faff{\mg}$, respectively). Indeed, one readily
verifies that ($\cN$ is an irrelevant normalisation constant)  
\beq\label{to-grav}
\lan a,b,\g_{ij} | \la \ran\ran_C= \ \cN \d^{ab} \d_{ij} \ \frac{S_{\g 
    \la}}{\sqrt{S_{\g 0}}}.  
\eeq
In the present context, we are dealing with a matrix model whose
elementary degrees of freedom are D$0$-branes (the D$0$-brane enters
the quantum-algebraic construction as \linebreak the trivial \rep of
$\tx{REA}_q(\mg)$, on which $M^{\La(0)} = \1_{d_\La}$), hence it seems
only natural to consider couplings normalised relative the reference
D$0$-brane,    
\beq\label{0rel-coupl}
\frac{\lan a,a, \g_{ij} | \la \ran\ran_C}{\lan b,b, \g_{kk} | 0
  \ran\ran_C} = \d_{ij} \ \frac{S_{\g \la}}{S_{\g 0}}.   
\eeq
From direct comparison between \refeq{explicit-f} and
\refeq{0rel-coupl}, we then draw the following  
\begin{conc}
\textit{The $\uqg$-tensor operators $M^{\La(\la)}, \ \La \in
  \faff{\mg}$, obtained from elementary solutions to the Reflection
  Equation through the Boundary-Weight Fusion \eqref{BWF} and
  composing a physically distinguished basis of the algebra of
  functions on the untwisted maximally symmetric WZW D-brane labeled
  by the weight $\la \in \faff{\mg}$, encode the complete information
  on bulk graviton $| a,b,\La^+_{ij} \ran$ couplings to the Cardy
  boundary state $| \la \ran\ran_C$ representing the D-brane, relative
  an elementary D$0$-brane, as expressed by the identity:  
\beq
\frac{\lan a,a,\La^+_{ij} | \la \ran\ran_C}{\lan b,b,\La^+_{kk} | 0
  \ran\ran_C} = \tr_q^{(\la)} (M^{\La}_{ij}). 
\eeq}   
\end{conc}
We emphasise that it is not just the numerical values of the couplings
but also \linebreak their structure, diagonal in the bulk \rep indices, that
can be read off from \refeq{explicit-f}. The D$0$-brane data (e.g. the
D$0$-brane tension), on the other hand, have to be supplemented
independently of the algebra\footnote{Actually, on the level of
  bulk-boundary couplings, the only piece of data that cannot be
  retrieved from the algebra is $S_{00}$. Indeed, we have \cite{cftbook}
\beqa
S_{\La^+ \la} &=& S_{0 \La} \cdot \chi_{\la} \lb \frac{2 \pi i (\La +  
\rho)}{\k + \gcheck(\mg)} \rb = S_{00} \cdot \chi_{\La} \lb \frac{2
  \pi i \rho}{\k + \gcheck(\mg)} \rb \cdot \chi_{\la} \lb \frac{2 \pi
  i (\La + \rho)}{\k + \gcheck(\mg)} \rb = S_{00} \cdot \tr_q^{(\La)}
\lb \tr_q^{(\la)} M^{\La} \rb, \non
\eeqa
where we have first used the symmetries of the modular $S$-matrix:
$S_{\la^+ \mu} = S_{\la \mu^+}$ and $S_{\la \mu} = S_{\mu \la}$, and
later reiterated the first equality.}.

\section{Conclusions}

In the present paper, we have discussed several application of the
algebraic RE fusion to the description of physics of untwisted
maximally symmetric WZW D-branes. The Bound-State Fusion has been shown 
to lead to the appropriate choice of realisations of the REA and to
give a nice picture of higher-dimensional quantum D-branes as
condensates of the elementary quantum D$0$-branes, reproducing - in
the semiclassical approximation - precisely the fuzzy condensation
scenario derived from stringy perturbation theory in
\cite{fuzzy-ars}. It also seems to offer some insight into the
structure of gauge fluctuations of the non-commutative geometry
defined by the quantised function algebra $\rea{\mg}$: being a
multiplicative perturbation of this background geometry, the gauge 
fluctuations are strongly reminiscent of the nonperturbative
Wilson-loop operators of Bachas and Gaberdiel \cite{bach}, which - in
turn - fits in well with earlier findings on the r\^ole of (open)
Wilson lines in gauge field theories on noncommutative geometries 
\cite{wilson}. The Bulk-Weight Fusion, on the other hand, has been
demonstrated to encode a fairly complete information on the
gravitational D-brane couplings. Both are amazingly simple and follow 
straightforwardly from the structure of \linebreak the RE.  

In spite of the progress, signified by our results, in formulating 
a compact description of quantum WZW geometry and elucidating the
quantum-group structure of the associated BCFT, a lot more still needs
to be understood in this context. We hope to return to \linebreak these riddles
soon. 

\acknowledgments  
The authors would like to thank the organisers of the 2004 ESI
Workshop on "String theory on non-compact and time-dependent
backgrounds" where part of this work was done. Our gratitude also goes
to the Theoretical Physics Group at King's College London and the
String Theory Group at Queen Mary London, and in particular to Thomas
Quella, Sanjaye Ramgoolam, Andreas Recknagel, Sylvain Ribault and
Steven Thomas, for their sustained interest in the project and 
stimulating discussions. R.R.S. is thankful to Krzysztof Gaw\c{e}dzki
for many illuminating discussions, and - in particular - for
explaining to him the Poisson structure behind the algebras of
interest and for drawing his attention to the papers expounding the
r\^ole it plays in the canonical description of the (boundary) WZW
model. Both authors gratefully acknowledge useful remarks from the
referee.   

\appendix 
\section{Quantum-group covariance}

Below, we consider $\uqg$-covariance properties of the various
generalised reflection matrices introduced in the main text. In
particular, we give a simple proof of \refeq{comm-proof} and
\refeq{si-tr}, essentially repeating the original one from
\cite{qalg}. First, we show, for $\cM_{12} := \cR_{21} \cR_{12}$,   
\beq\label{comm}
\Delta (u) \cM_{12} = \Delta(u) \cR_{21} \cR_{12} = \cR_{21}
\Delta^{cop}(u) \cR_{12} = \cR_{21} \cR_{12} \Delta(u) = \cM_{12}
\Delta (u),  
\eeq
where we have invoked the twisting property of $\cR$
\cite{majbook,chp}:  
\beq\label{R-cop-twist}
\D^{cop} (u) = \cR \D (u) \cR^{-1}, \qquad \D^{cop} (u) := u_2 \ox
u_1. 
\eeq
Using Hopf-algebra identities for the coproduct and the antipode of
$\uqg$ (i.e. taking \refeq{comm} with both sides of the identity
extended by $S u_0 \ox I$ from the left and by $I \ox S u_3$ from \linebreak
the right, and - upon contracting and then multiplying the tensor factors
in the pairs of spaces $(0,1)$ and $(2,3)$ - representing both sides
on $V_\La \ox V_\la$), we turn \refeq{comm} into \refeq{comm-proof},
or    
\beq\label{univtens}
\pi_\la (u_1) M^{\La(\la)} \pi_\la (Su_2) = \pi_\La (Su_1) M^{\La(\la)} 
\pi_\La (u_2),
\eeq
for any $u \in \uqg$.

In order to prove \refeq{si-tr}, we recall the definition of the quantum 
trace: $\tr_q (x) := \tr (x \v)$, where $\v := (S \ox \id)(\cR_{21})$ is 
the distinguished invertible ($\v^{-1} \equiv S \v$) element of $\uqg$ 
satisfying $S^2 u = \v u \v^{-1}$ for any $u \in \uqg$ \cite{majbook}. 
This, together with \refeq{comm-proof}, immediately implies 
\beqa
\pi_\La (Su_1) \tr_q^{(\la)} ( M^\La ) \pi_\La (u_2) = \tr_q^{(\la)}( 
u_1 M^\La Su_2) = \tr_q^{(\la)} ( M^\La S u_2 \v u_1) = \eps(u) 
\tr_q^{(\la)} ( M^\La ),
\eeqa
or, equivalently,
\beq
[\pi_\La(u), \tr_q^{(\la)}(M^\La)] = 0.
\eeq

Last, we may verify the tensorial character of \refeq{BSF-0}, on which
our physical interpretation of the BSF has been based. Our proof is in
fact a slight variation of the trick used above. We begin by defining
the operator $\cM_{123} = \cR_{32} \cR_{23} \cR_{21} \cR_{31} \cR_{13}
\cR_{12}$ such that $M^{\La,\La_B \x \la}_{1 \ 2 \ox 3} \equiv
(\pi_\La \ox \pi_{\La_B} \ox \pi_\la)(\cM_{123})$. Using
\refeq{R-cop-twist} again, we then obtain 
\beq
[(\D \ox \id) \ox \D](u) \cM_{123} = \cM_{123} [(\D \ox \id) \ox
  \D](u) 
\eeq 
and hence
\beq
(Su_1 \ox I \ox I) \cM_{123} (u_2 \ox I \ox I) = (I \ox u_1 \ox u_2)
\cM_{123} (I \ox Su_4 \ox Su_3).  
\eeq
The latter formula ultimately turns into an appropriate analogon of
\refeq{univtens},
\beq\label{tensop}
\pi_\La(Su_1) M^{\La,\La_B \x \la} \pi_\La(u_2) = \pi_{\La_B \ox
  \la}(u_1) M^{\La,\La_B \x \la} \pi_{\La_B \ox \la}(Su_2),
\eeq
once we invoke one of the fundamental properties of a Hopf algebra
\cite{majbook}, $\D \circ S = (S \ox S) \circ \D^{cop}$, and use the
standard definition of a tensor-product \rep of a coalgebra,
$\pi_{\La_1 \ox \La_2} := (\pi_{\La_1} \ox \pi_{\La_2}) \circ \D$.

\end{document}